\documentclass[prd,aps, tightenlines, preprintnumbers, showpacs, nofootinbib,superscriptaddress,notitlepage,11pt]{revtex4-1}

\usepackage{epsfig}
\usepackage{amsmath,amsthm,amssymb}
\usepackage{mathtools}
\usepackage{siunitx}

\newcommand{\beq}{\begin{eqnarray}}
\newcommand{\eeq}{\end{eqnarray}}

\newcommand{\nn}{\nonumber \\}

\newif\iffigure
\figuretrue

\begin{document}
\preprint{YITP-17-22}

\title{Gluon Tomography from Deeply Virtual Compton Scattering at Small-$x$}

\author{Yoshitaka Hatta}
\affiliation{Yukawa Institute for Theoretical Physics, Kyoto University, Kyoto 606-8502, Japan}

\author{Bo-Wen Xiao}
\affiliation{Key Laboratory of Quark and Lepton Physics (MOE) and Institute
of Particle Physics, Central China Normal University, Wuhan 430079, China}

\author{Feng Yuan}
\affiliation{Nuclear Science Division, Lawrence Berkeley National
Laboratory, Berkeley, CA 94720, USA}

\begin{abstract}
We present a full evaluation of the deeply virtual Compton scattering (DVCS) cross section in the dipole framework in the small-$x$ region. The result features the $\cos \phi$ and $\cos 2\phi$ azimuthal angular correlations which have been missing in previous  studies based on the dipole model. In particular, the $\cos 2\phi$ term is generated by the elliptic gluon Wigner distribution whose measurement at the planned electron-ion collider (EIC) provides an important 
information about the gluon tomography at small-$x$. We also show the consistency with the standard collinear factorization approach based on the quark and gluon generalized parton distributions (GPDs). 
\end{abstract}
\pacs{24.85.+p, 12.38.Bx, 12.39.St}
\maketitle

\section{Introduction}
The deeply virtual Compton scattering (DVCS) is one of the most important channels to study 
the partonic structure of nucleon, in particular, to unveil the orbital angular momentum information
for the quarks and gluons~\cite{Ji:1996ek,Mueller:1998fv,Ji:1996nm,Radyushkin:1997ki}. 
It has attracted tremendous interests from both
theory and experimental sides~\cite{Goeke:2001tz,Diehl:2003ny,Belitsky:2005qn,Boffi:2007yc,Boer:2011fh,Accardi:2012qut}. 
Experimentally, it is a simple high energy scattering
process, and is a major emphasis in the current and future lepton-nucleon
collision facilities~\cite{Boer:2011fh,Accardi:2012qut}. 
Among the observables in DVCS, it has been predicted that there exists a 
$\cos 2\phi$ azimuthal correlation due to the so-called helicity-flip gluon generalized parton distributions  
(GPDs)~\cite{Diehl:1997bu,Hoodbhoy:1998vm,Belitsky:2000jk,Diehl:2001pm,Belitsky:2001ns}. 
In this paper, we investigate this physics in the small-$x$ dipole formalism, which is also known as the color-glass condensate 
(CGC) formalism~\cite{McLerran:1993ni}. We will show that the $\cos 2\phi$ correlation in DVCS 
provides a unique opportunity to test the CGC prediction, and at the same time provides crucial
information on the gluon tomography at small-$x$, in particular, that associated
with the so-called elliptic gluon distribution~\cite{Hatta:2016dxp,Hagiwara:2016kam,Zhou:2016rnt,Hagiwara:2017ofm}. 

In the small-$x$ dipole factorization approach, the DVCS amplitude can be 
schematically calculated as~\cite{Donnachie:2000px,Bartels:2003yj,Favart:2003cu,Kowalski:2006hc}
\begin{equation}
{\cal A}_{DVCS}\sim \int d^2b_\perp e^{ib_\perp \cdot \Delta_\perp} \int dzd^2r_\perp \Psi_{\gamma^*}(z,r_\perp)\Psi_{\gamma}^*(z,r_\perp)
{\cal T}(b_\perp, r_\perp) \ ,
\end{equation}
where $\Psi$ and $\Psi^*$ are the wave functions for the incoming 
virtual photon and outgoing real photon, respectively. The physics behind this factorization can be 
understood as illustrated in Fig.~\ref{fac0}, where the virtual photon fluctuates into
a quark-antiquark pair to form a color-dipole. The latter scatters
on the nucleon target and merges into a real photon in the final state,
whereas the nucleon recoils with momentum transfer $\Delta$. 
The wave functions depend on the momentum fraction of the photon
carried by the quark $z$ and the
dipole size $r_\perp$. For sufficiently hard scatterings, they are perturbatively calculable. In the DVCS
amplitude, ${\cal T}$ describe the elastic scattering of the dipole with the 
nucleon target. This is different from the inclusive deep inelastic
scattering, which depends on the inelastic scattering amplitude. 
The elastic scattering amplitude can be written as
\begin{equation}
{\cal T}=1-S \ ,
\end{equation}
where $S$ represents the dipole S-matrix (defined below). 
In the previous calculations of DVCS in the CGC formalism, 
the main focus is on the azimuthally symmetric cross section 
in which the photon helicity is conserved. In order
to obtain the azimuthal $\cos 2\phi$ correlation, we need
to carry out the calculation on the helicity-flip amplitude. We perform our
calculations in both coordinate space and momentum space
and check their consistency.

\iffigure
\begin{figure}[t]
\begin{center}
\includegraphics[width=8cm]{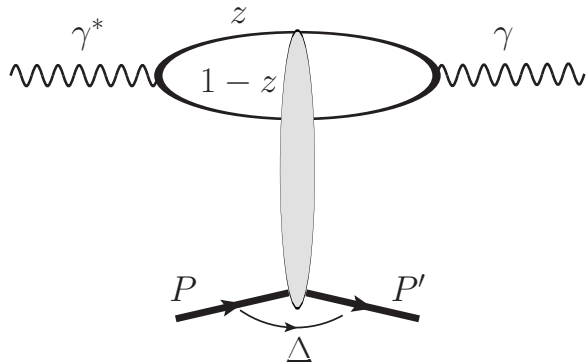}
\end{center}
\caption[*]{Deeply virtual Compton Scattering $\gamma^*p\to \gamma p$ in the small-$x$ limit.
}
\label{fac0}
\end{figure}
\fi
 
An important aspect of our calculations is the 
comparison with the collinear factorization results. The key observation is the connection 
between the gluon GPDs at small-$x$ and the dipole scattering amplitude. 
For the $\cos 2\phi$ azimuthal correlation in the DVCS
process, we show that the helicity-flip amplitude calculated from the elliptic gluon distribution reduces, in the collinear limit, to 
that from the helicity-flip gluon GPD in the collinear framework.
Meanwhile, for the azimuthally symmetric cross section,
the dipole formalism leads to divergence in the collinear  
limit. This can be interpreted as the $\mathcal{O}(\alpha_s)$ contribution to the quark GPD in the collinear framework, according to the relation between the quark GPD
and the gluon GPD at small-$x$. These results establish a complete 
consistency between the CGC
formalism and the collinear factorization framework.

The rest of the paper is organized as follows. In Sec.~II, we derive the small-$x$
gluon GPDs in the CGC formalism. The two GPDs are expressed in terms of the
gluon Wigner distributions. In particular, the so-called elliptic gluon Wigner
distribution will contribute to the helicity-flip gluon GPD. 
In Sec.~III, we calculate the DVCS amplitude in the dipole framework in coordinate space  and derive the $\cos 2\phi$ correlation. 
In Sec.~IV, we perform the calculations in  momentum space
and demonstrate the consistency with the coordinate space derivations
in Sec.~III. The comparisons to the collinear factorization
results will be made in Secs.~III and IV and Appendix A. 
In Sec.~V, we compute the contribution from the longitudinally polarized virtual photon and find the $\cos \phi$ correlation.  
Finally, we summarize our paper in Sec.~VI.

\section{The dipole S-matrix and the gluon GPD}
\label{2}

In this section, we introduce the basic ingredient to calculate the DVCS amplitude at small-$x$, namely, the dipole S-matrix. We shall clarify the relation between  the gluon GPDs and the dipole S-matrix, and show that the latter provides an efficient description of the DVCS amplitude which is free of collinear divergences.

\iffigure
\begin{figure}[t]
\begin{center}
\includegraphics[width=17cm]{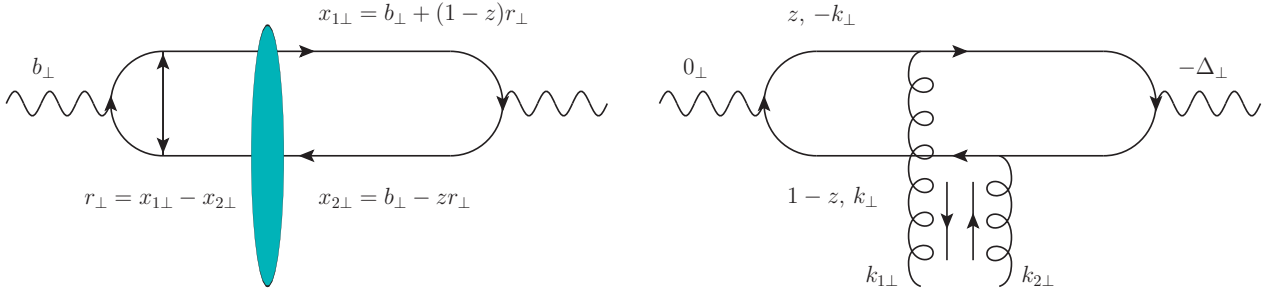}
\end{center}
\caption[*]{Left diagram: DVCS amplitude in transverse coordinate space; Right diagram: DVCS amplitude in momentum space.}
\label{cm}
\end{figure}
\fi

In the dipole framework, the DVCS amplitude is represented by the diagram in Fig.~\ref{cm} in coordinate space (left) and in 
momentum space (right). We work in a frame in which 
the virtual photon and the proton are collinear,  with the proton moving fast in the positive $z$-direction. In coordinate space, we fix the transverse coordinates of the quark and anti-quark to be $x_{1\perp}=b_\perp + (1-z)r_\perp$ and $x_{2\perp}=b_\perp -zr_\perp$, respectively, with $z$ defined as the longitudinal 
momentum fraction of the quark with respect to the incoming virtual photon. The `center-of-mass' of the $q\bar q$ system 
coincides with the virtual photon coordinate $zx_{1\perp} +(1-z) x_{2\perp}=b_\perp$. The size of the $q\bar q$ system is $r_\perp = x_{1\perp}- x_{2\perp}$. In this setup, the forward S-matrix for the $q\bar q$ pair scattering off the target reads 
 \begin{eqnarray}
S_x\bigl(b_\perp +(1-z) r_\perp, b_\perp- zr_\perp\bigr) \equiv \left\langle \frac{1}{N_c} \textrm{Tr} \left[U(b_\perp+(1-z)r_\perp) U^\dagger (b_\perp-zr_\perp)\right] \right\rangle_x\,, \label{s}
\end{eqnarray}
where $x$ is the relevant momentum fraction of gluons in the target. In DVCS and in the small-$x$ limit, it is related to the Bjorken variable $x_{Bj}$ as $x\approx \frac{x_{Bj}}{2}$, which is also the same as the skewness parameter $\xi$ (defined below). $U$ is the Wilson line
\beq
U(x_\perp)=P\exp\left(-ig\int_{-\infty}^\infty dx^- A^+(x^-,x_\perp)\right),
\eeq
which represents the eikonal propagation of the quark. The brackets $\langle...\rangle$ denote the off-forward proton matrix element $\frac{\langle p'|...|p\rangle}{\langle p|p\rangle}$ with $p'=p+ \Delta$. 
In momentum space, we define
\beq
{\cal F}_x(\tilde{q}_\perp,\Delta_\perp,z) &\equiv& \int \frac{d^2r_\perp d^2b_\perp}{(2\pi)^4} e^{i\Delta_\perp \cdot b_\perp  + i \tilde{q}_\perp \cdot r_\perp} S_x\bigl( b_\perp +(1-z) r_\perp, b_\perp- zr_\perp\bigr) \nn
 &=& \int \frac{d^2r_\perp d^2b'_\perp}{(2\pi)^4} e^{i\Delta_\perp \cdot b'_\perp  + i\tilde{q}_\perp \cdot r_\perp} 
 e^{-i\delta_\perp\cdot r_\perp}
 S_x\left(b'_\perp +\frac{r_\perp}{2}, b'_\perp- \frac{r_\perp}{2}\right)  \nn
&=& F_x(q_\perp\equiv\tilde{q}_\perp-\delta_\perp,\Delta_\perp), \label{com}
\eeq
 where $\delta_\perp \equiv \frac{1-2z}{2}\Delta_\perp$ and 
\beq
 F_x(q_\perp,\Delta_\perp) &=&  \int \frac{ d^2r_\perp d^2b_\perp }{(2\pi)^4} e^{ib_\perp \cdot \Delta_\perp + ir_\perp \cdot q_\perp}  S_x\left(b_\perp +\frac{r_\perp}{2}, b_\perp- \frac{r_\perp}{2}\right) . \label{f}
\eeq
In momentum space, we can also write ${\cal F}_x=\frac{1}{(2\pi)^4}\int d^2x_{1\perp} d^2x_{2\perp} e^{ik_{1\perp}\cdot x_{1\perp}-ik_{2\perp}\cdot x_{2\perp}} S_x(x_{1\perp}, x_{2\perp})$ with $k_{1\perp}\equiv \tilde{q}_\perp +z\Delta_\perp $ and $k_{2\perp}\equiv \tilde{q}_\perp -(1-z)\Delta_\perp$ conjugate to $x_{1\perp}$ and $x_{2\perp}$, respectively. The directions of transverse momenta flow of exchanged gluons are labeled in Fig.~\ref{cm}. Following \cite{Hatta:2016dxp}, we decompose $F$ into the angular independent and `elliptic' parts
\beq
 F_x(q_\perp,\Delta_\perp) = F_0(|q_\perp|,|\Delta_\perp|) + 2\cos 2(\phi_{q_\perp}-\phi_{\Delta_\perp})F_\epsilon (|q_\perp|,|\Delta_\perp|)+\cdots .
\eeq
Below $F_\epsilon$ will be referred to as the elliptic gluon distribution. It is at most a few percent in magnitude compared to $F_0$, but has  very different functional dependencies on $x$ and $q_\perp$ \cite{Hagiwara:2016kam}. It can thus lead to distinct experimental signatures \cite{Hatta:2016dxp,Zhou:2016rnt,Hagiwara:2017ofm}. One of the main goals of this paper is to clarify the role of $F_\epsilon$ in DVCS.

Comments are in order regarding the phase factor $e^{-i\delta_\perp \cdot r_\perp}$ in (\ref{com}).  In Ref.~\cite{Kowalski:2006hc}, the authors introduced a phase factor in the DVCS  amplitude in the $b_\perp$-space
\beq
\frac{d^2\sigma}{d^2b_\perp}= 2(1-S(b_\perp,r_\perp)) \to \frac{d^2\sigma}{d^2b_\perp} e^{-i(1-z)\Delta_\perp\cdot r_\perp}, \label{ko}
\eeq
 and this prescription has been used in many subsequent works \cite{Goncalves:2007sa,Berger:2012wx,Rezaeian:2013tka,Goncalves:2014wna,Xie:2016ino}.  It is motivated by the explicit perturbative analysis in  \cite{Bartels:2003yj} that such a phase factor arises in nonforward amplitudes $\Delta_\perp\neq 0$. However, the result of \cite{Bartels:2003yj} has been misinterpreted. To see the problem, note that (\ref{ko}) is not invariant under the combined transformation $z\to 1-z$ and $r_\perp \to -r_\perp$.
This transformation interchanges quark  and antiquark, and has been emphasized in  \cite{Bartels:2003yj} as the exact symmetry of the dipole formalism. The phase  factor discussed in  \cite{Bartels:2003yj}  ensures that the effective transverse coordinates of the quark and antiquark is $b_\perp + (1-z)r_\perp$ and $b_\perp-zr_\perp$, respectively, and this has been taken into account in (\ref{s}). Eq.~(\ref{com}) then shows that the correct phase factor should be $e^{-i\delta_\perp \cdot r_\perp} = e^{-i\frac{1-2z}{2}\Delta_\perp \cdot r_\perp}$ which is by itself invariant under the transformation $z\to 1-z$ and $r_\perp \to -r_\perp$.  
As a nontrivial crosscheck, in Section~\ref{mom} we compute the DVCS amplitude in the momentum space and find the equivalent of this phase factor. 
We then show in Section~V that this phase factor plays an important role in DVCS processes involving the longitudinally polarized virtual photon. We remark in passing that no phase factor is needed in the case of diffractive dijet production \cite{Hatta:2016dxp}, though the process looks rather similar to DVCS.

\subsection{Relation to GPD at small-$x$}

Let us point out the relation between $F_0$ and $F_\epsilon$ introduced above and the gluon GPDs which are defined as
\beq
&&\frac{1}{P^+} \int \frac{d\zeta^-}{2\pi}e^{ix P^+\zeta^-} \langle p'|F^{+i}(-\zeta/2)F^{+j}(\zeta/2)|p\rangle \nn
&& \qquad \quad =\frac{\delta^{ij}}{2}xH_g(x,\Delta_\perp)+ \frac{xE_{Tg}(x,\Delta_\perp)}{2M^2}\left( \Delta^i_\perp \Delta^j_\perp- \frac{\delta^{ij}\Delta_\perp^2}{2}\right)+\cdots, \label{left}
\eeq
where $M$ is the proton mass and $P=\frac{p+p'}{2}$. Our convention for the gluon GPDs is such that $H_g(x,\Delta_\perp\to 0)=G(x)$ (the unpolarized gluon PDF) in the forward limit. The helicity-flip gluon GPD $E_{Tg}$ is also called the gluon transversity GPD, and the above normalization coincides with that of \cite{Hoodbhoy:1998vm}.\footnote{It differs from the normalization in \cite{Diehl:2003ny} by a factor $-2x$.}   We suppress the dependence of GPDs on the skewness parameter  
$\xi = {(p^+ - p'^+)/( p^+ + p'^+)}$. 
Unless otherwise specified, it is understood that 
$H_g(x,\Delta_\perp)\equiv H_g(x,\xi=x,\Delta_\perp)$ and $E_{Tg}(x,\Delta_\perp)\equiv E_{Tg}(x,\xi=x,\Delta_\perp)$. This is  because the imaginary part of the DVCS amplitude, which we assume to be  dominant at small-$x$, probes GPDs at $\xi=x$ to leading order. 
It is also known that,  
for the gluon GPDs at small-$x$, this dependence has been found to be very mild, see for example the discussions in Ref.~\cite{Diehl:2003ny}, which is consistent with the color-dipole formalism. The leading contribution of the 
S-matrix in the dipole formalism does not differentiate the dependence on $x$ and $\xi$.

At small-$x$, the left hand side of (\ref{left}) can be approximately written as \cite{Hatta:2016dxp},
\beq
&& \frac{1}{P^+} \int \frac{d\zeta^-}{2\pi}e^{ix P^+\zeta^-} \langle p'|F^{+i}F^{+j}|p\rangle
\approx  \frac{2N_c}{\alpha_s} \int d^2q_\perp   \left(q_\perp^i -\frac{\Delta_\perp^i}{2}\right)\left(q_\perp^j+\frac{\Delta_\perp^j}{2}\right)  F(q_\perp,\Delta_\perp) \nn 
&& \qquad= \frac{2N_c}{\alpha_s}\int d^2q_\perp q_\perp^i q_\perp^j \left[ F_0(|q_\perp|,|\Delta_\perp|) + 2\left(\frac{2(\vec{q}_\perp \cdot \vec{\Delta}_\perp)^2}{q_\perp^2\Delta_\perp^2}-1 \right) F_\epsilon(|q_\perp|,|\Delta_\perp|) \right] \nn
&& \qquad = \frac{2N_c}{\alpha_s} \left( \frac{\delta^{ij}}{2} \int d^2 q_\perp q_\perp^2 F_0  + \frac{1}{\Delta_\perp^2}\left(\Delta^i_\perp \Delta^j_\perp -\frac{\delta^{ij}\Delta_\perp^2}{2}\right)  \int d^2 q_\perp q_\perp^2 F_\epsilon\right)\,,
\eeq
where we used the fact that $\int d^2q_\perp F(q_\perp,\Delta_\perp)=0$ for $\Delta_\perp \neq 0$. 
We thus obtain important relations between the gluon GPDs and the small-$x$ dipole distributions as follows
\beq
xH_g(x,\Delta_\perp) &=& \frac{2N_c}{\alpha_s}  \int d^2 q_\perp q_\perp^2 F_0  \,, \label{f0} \\
xE_{Tg}(x,\Delta_\perp) &=&   \frac{4N_c M^2}{\alpha_s \Delta_\perp^2} \int d^2 q_\perp q_\perp^2 F_\epsilon\,. \label{fe}
\eeq
These formulas will be used below to check the consistency with the collinear approach. The physical interpretation of the gluon GPDs and the above relations becomes manifest 
in the following computations of DVCS amplitudes.

\section{DVCS Amplitude and $\cos 2\phi$ Azimuthal Angular Correlation}

The differential cross section for DVCS can be written as
\begin{eqnarray}
\frac{d\sigma(ep\to e'\gamma p')}{dx_{Bj}dQ^2d^2\Delta_\perp}=\frac{\alpha_{em}^3 x_{Bj}y^2}{4\pi Q^4}\frac{L_{\mu\nu}{\cal M}^{\mu\nu}}{Q^4} \ , \label{start}
\end{eqnarray}
where ${ L}_{\mu\nu}$ is the lepton tenor and ${\cal M}^{\mu\nu}$ is the hadronic
tensor. 
We use vectors $l$ and $l'$ for the initial and final state lepton momenta, $p$ and $p'=p+\Delta$ for
the initial and final state proton momenta, respectively. The incoming virtual photon has momentum 
$q = l - l'$ with virtuality $q^2 =-Q^2$ with vanishing transverse momentum. We use the standard variables $x_{Bj}= {Q^2/ (2q\cdot p)}$, $y= {q\cdot p /(l\cdot p)}$. 
$t=-\Delta^2_\perp$ and  $W^2=(q+p)^2 \approx  Q^2/x_{Bj}$. In (\ref{start}), we only take into account  the DVCS process and neglect the Bethe-Heitler contribution.  
In fixed-target experiments such as at COMPASS where $Q$ is at most a few GeV or less at small-$x$, the cross section is dominated by the Bethe-Heitler contribution.  In collider experiments such as at HERA and the EIC, especially at large center-of-mass energies and small-$x$, there exist regions in kinematic variables where the  cross section is dominated by the DVCS process \cite{Aktas:2005ty,Aschenauer:2013hhw}. We focus on the latter situation throughout this paper.

The hadronic tensor can be decomposed as 
\beq
{\cal M}^{\mu\nu}= {\cal M}_{TT}^{\mu\nu} + {\cal M}^{\mu\nu}_{TL}+{\cal M}_{LL}^{\mu\nu}\ ,
\eeq
where the subscripts $T$ and $L$ (transverse and longitudinal) denote the polarizations of the virtual photon in the amplitude and  complex-conjugate amplitudes. (The outgoing real photon is always transversely polarized.) 
In this and the next sections, we will focus on ${\cal M}_{TT}$. The longitudinally polarized case will be treated in Section \ref{long}. 
In the present frame, 
the lepton tensor can be decomposed into, for $\mu,\nu$ transverse, 
\begin{eqnarray}
L_{\mu\nu}&=&2(l_\mu l'_\nu+l_\nu l'_\mu-g_{\mu\nu}l\cdot l')\nonumber\\
&=&\frac{2Q^2}{y^2}\left[\left(1-y+\frac{y^2}{2}\right)g_{\perp\mu\nu}+(1-y)\hat h_{\perp\mu\nu}\right] \ ,
\end{eqnarray}
where $g_{\perp}^{\mu\nu}=-g^{\mu\nu}+(\hat{p}^\mu \hat{n}^\nu+\hat{p}^\nu \hat{n}^\mu)/\hat{p}\cdot \hat{n}$
and $\hat h_\perp^{\mu\nu}=\frac{2l_\perp^\mu l_\perp^\nu}{l_\perp^2}-g_\perp^{\mu\nu}$.
$\hat{p}$ and $\hat{n}$ are two light-like vectors: $\hat{p}^2=\hat{n}^2=0$ and $\hat{p}\cdot \hat{n}=1$.
Here $l_\perp=l'_\perp$ represents the transverse momentum of the lepton. It satisfies the relation  $l_\perp^2 = \frac{1-y}{y^2}Q^2$. 
The hadronic tensor is calculated from the amplitude squared of $\gamma^*+p\to \gamma+p'$,
\begin{eqnarray}
{\cal M}^{\mu\nu}_{TT}=W^4g_{\perp\alpha\beta}{\cal A}^{\mu\alpha}_T\left({\cal A}^{\nu\beta}_T\right)^* \ ,
\end{eqnarray}
where $\mu$, $\nu$ represent the (transverse) polarization indices for the incoming 
virtual photon, and $\alpha$, $\beta$ for the outgoing photon, respectively. We have defined ${\cal A}^{\mu\nu}$ as the imaginary part of the amplitude. The real part is subleading at small-$x$ and can be retrieved 
through the dispersion relation, if necessary. It is convenient to decompose the tensor indices as 
\begin{eqnarray}
{\cal A}_T^{\mu\nu}(\Delta_\perp)=g_\perp^{\mu\nu}{\cal A}_{0} (\Delta_\perp)+h_\perp^{\mu\nu}{\cal A}_{2}  (\Delta_\perp)\ ,
\label{co}
\end{eqnarray}
where $h_\perp^{\mu\nu}=\frac{2\Delta_\perp^\mu \Delta_\perp^\nu}{\Delta_\perp^2}-g_\perp^{\mu\nu}$.
The differential cross section then takes the form
\begin{eqnarray}
\frac{d\sigma_{TT}}{dx_BdQ^2d^2\Delta_\perp}=\frac{\alpha_{em}^3 }{\pi x_{Bj}Q^2}\left\{
\left(1-y+\frac{y^2}{2}\right)({\cal A}_0^2+{\cal A}_2^2)+(1-y) 2{\cal A}_0{\cal A}_2 \cos(2\phi_{\Delta l})  \right\}\ ,
\label{x}
\end{eqnarray}
where $\phi_{\Delta l}$ is the azimuthal angle of the final state photon with 
respect to the lepton plane.
The amplitudes ${\cal A}_{0,2}$ can be calculated from different 
projections of the tensor ${\cal A}_T^{\mu\nu}$. Alternatively, as noted in Refs.~\cite{Diehl:1997bu,Hoodbhoy:1998vm}, they can also be obtained from the 
helicity conserved and helicity-flip amplitudes as  
\begin{eqnarray}
\frac{1}{2}\sum_{\lambda}\mathcal{A}^{\lambda=\lambda^\prime}_T (\Delta_\perp) ={\cal A}_0\,, \qquad
 \frac{1}{2}\sum_{\lambda}\mathcal{A}^{\lambda\neq \lambda^\prime}_T (\Delta_\perp) =-{\cal A}_2\cos 2\phi_{\Delta_\perp}\,, \label{sum}
\end{eqnarray}
where $\lambda$ and $\lambda'$ represent the helicities of the incoming
and outgoing photons. 
${\cal A}_T^{\lambda\lambda'}\equiv \epsilon^\lambda_\mu {\cal A}_T^{\mu\nu}\epsilon^{*\lambda'}_\nu$ can be conveniently expressed in  coordinate space using the dipole S-matrix introduced in Section II
\begin{eqnarray}
\mathcal{A}^{\lambda, \lambda'}_T (\Delta_\perp) &=& 2 \int d^2 b_\perp e^{ib_\perp\cdot \Delta_\perp}  N_c\sum_q \int d^2 r_\perp\int_0^1 \frac{dz}{4\pi} \Psi^\lambda _{\gamma^{ \ast}} (z, r_\perp) \Psi^{\lambda' \ast}_{\gamma} (z, r_\perp)  \nonumber \\
 &&  \qquad \qquad \times \bigl(1-S(b_\perp +(1-z) r_\perp, b_\perp-zr_\perp)\bigr),  \label{As}
\end{eqnarray}
 where $\Psi$ is the photon wavefunction. For the incoming virtual photon, it is given by 
\begin{align}
\Psi^{T\,\lambda}_{\gamma^\ast\alpha\beta}(z,r_\perp)&=\frac{ie_q}{\pi}\epsilon_qK_1(\epsilon_q|r_\perp|)\begin{cases}\tfrac{r_\perp\cdot\epsilon^{(1)}_\perp}{|r_\perp|}[\delta_{\alpha+}\delta_{\beta+}z -\delta_{\alpha-}\delta_{\beta-}(1-z)], & \lambda=1,\\\tfrac{r_\perp\cdot\epsilon^{(2)}_\perp}{|r_\perp|}[\delta_{\alpha-}\delta_{\beta-}z-\delta_{\alpha+}\delta_{\beta+}(1-z)], & \lambda=2,\end{cases}\\
\Psi^L_{\gamma^\ast\alpha\beta}(z,r_\perp)&=\frac{e_q z(1-z)Q}{\pi} K_0(\epsilon_q|r_\perp|)\delta_{\alpha\beta},
\end{align}
where  $\alpha$ and $\beta$ are the quark and
antiquark helicities, $e_q$ is the electric charge of the quark (in units of $e$)  
and $\epsilon_q^2=z(1-z)Q^2$. The quark mass has been neglected. 
For the outgoing real photon, we have
\begin{align}
\Psi^{T\,\lambda}_{\gamma\alpha\beta}(z,r_\perp)&=e_q\frac{i}{\pi} \begin{cases}\tfrac{r_\perp\cdot\epsilon^{(1)}_\perp}{r_\perp^2}[\delta_{\alpha+}\delta_{\beta+}z-\delta_{\alpha-}\delta_{\beta-}(1-z)], & \lambda=1,\\ \tfrac{r_\perp\cdot\epsilon^{(2)}_\perp}{r_\perp^2}[\delta_{\alpha-}\delta_{\beta-}z-\delta_{\alpha+}\delta_{\beta+}(1-z)], & \lambda=2.\end{cases} 
\end{align}

\subsection{Helicity Conserved Amplitude}
From (\ref{sum}) and (\ref{As}), we immediately find 
\begin{eqnarray}
{\cal A}_0&=&\frac{1}{2}\sum_{\lambda}\mathcal{A}^{\lambda=\lambda^\prime}_T (\Delta_\perp) \nonumber\\
&=& -\sum_q\frac{e_q^2 N_c}{\pi}  \int_0^1 dz \left[z^2+(1-z)^2\right]  \int \frac{d^2 r_\perp}{r_\perp}  \epsilon_qK_1(\epsilon_q r_\perp)\int d^2q_\perp e^{-iq_\perp \cdot r_\perp}  {\cal F}(q_\perp, \Delta_\perp,z) \, , \nn
  &=& -\sum_q \frac{e_q^2 N_c}{\pi} \int_0^1 dz \left[z^2+(1-z)^2\right]  \int \frac{d^2 r_\perp}{r_\perp}  \epsilon_qK_1(\epsilon_q r_\perp)  \nn 
&& \qquad \times \int d^2 q_\perp e^{-iq_\perp \cdot r_\perp-i\delta_\perp \cdot r_\perp}\Bigl(F_0(|q_\perp|,|\Delta_\perp|) +2 \cos 2(\phi_{q_\perp}-\phi_{\Delta_\perp}) F_\epsilon(|q_\perp|,|\Delta_\perp|) \Bigr) \nn 
&=& -\sum_q 2e_q^2 N_c \int_0^1 dz \left[z^2+(1-z)^2\right]  \int_0^\infty d r_\perp  \epsilon_qK_1(\epsilon_q r_\perp)  \nn && \times 
 \int d^2q_\perp \Bigl( J_0(|q_\perp+\delta_\perp|r_\perp)F_0(|q_\perp|,|\Delta_\perp|) + 2J_2(\delta_\perp r_\perp)J_2(q_\perp r_\perp) F_\epsilon(|q_\perp|,|\Delta_\perp|) \Bigr). \label{from}
\end{eqnarray}
Let us first consider the $F_0$ term in the last line. The $r_\perp$-integral looks divergent at first sight, since $\int_0^\infty d r_\perp  \epsilon_qK_1(\epsilon_q r_\perp) J_{0}(q_\perp r_\perp)$ is logarithmically divergent at $r_\perp =0$. However, this divergence is not physical and it can be removed easily. Using the fact that $\int d^2q_\perp   F(q_\perp, \Delta_\perp)=0$, we obtain a convergent result 
\begin{equation} 
\int_0^\infty d r_\perp  \epsilon_qK_1(\epsilon_q r_\perp) \left[J_{0}(|q_\perp+\delta_\perp| r_\perp)-1\right]=-\frac{1}{2} \ln \left[1+\frac{(q_\perp+\delta_\perp)^2}{\epsilon_q^2}\right].
\end{equation}
The $r_\perp$-integral in the $F_\epsilon$ term can also be done analytically in terms of the Appell function (see the formula 6.578-2 in \cite{grad}). We may however neglect this term as a higher order effect $J_2(\delta_\perp r_\perp) \sim \delta^2_\perp$  and obtain
\begin{eqnarray}
{\cal A}_0(\Delta_\perp)
  \approx \sum_q e_q^2 N_c \int_0^1 dz \left[z^2+(1-z)^2\right] \int d^2 q_\perp \ln \left[1+\frac{(q_\perp+\delta_\perp)^2}{z(1-z)Q^2}\right]  F_0 (|q_\perp|, |\Delta_\perp|)\ .
\end{eqnarray}

If one wishes to make contact with the collinear approach, one can expand the logarithm to linear order in $(q_\perp+\delta_\perp)^2$ and find that only the $q_\perp^2$ term survives after the $d^2q_\perp$ integration. Thus one recovers the GPD $xH_g (x, \Delta_\perp)$, see (\ref{f0}).  However, the prefactor is divergent due to the poles at $z=0, 1$. In order to isolate this divergence, one needs to return to the last line of  (\ref{from}) and employ the dimensional regularization in coordinate space as discussed in the appendix of Ref.~\cite{Mueller:2013wwa}. That is, in the  $\overline{\textrm{MS}}$ scheme, one can modify the $r_\perp$-integral as\footnote{This is equivalent to the dimensional regularization with $d=2-2\varepsilon$ in the momentum space.}  
\begin{equation}
\int \frac{d^2 r_\perp}{(2\pi)^2} \quad \to \quad \bar\mu ^{2 \varepsilon} (4\pi e^{-\gamma_E})^{\varepsilon}  \int \frac{d^{2+2\varepsilon} r_\perp}{(2\pi)^{2+2\varepsilon}}, \quad \textrm{with} \quad \bar\mu ^2=\frac{\mu^2}{4e^{-2\gamma_E}}.  \label{dr} 
\end{equation}

Expanding $J_0 (q_\perp r_\perp) =1-\frac{1}{4} q_\perp^2 r_\perp^2 +\cdots$ and  keeping only the second term which is the leading twist contribution, we find 
\begin{eqnarray}
-\frac{1}{4}\int_0^1 dz \left[z^2+(1-z)^2\right] \int \frac{d^2 r_\perp}{2\pi r_\perp}  \epsilon_qK_1(\epsilon_q r_\perp) r_\perp^2 &\to& -\frac{1}{Q^2}\left(\frac{Q^2 e^{-\gamma_E}}{\mu^2}\right)^{-\varepsilon} \frac{\Gamma(2-\varepsilon)\Gamma (2+\varepsilon)\Gamma(-\varepsilon)}{\Gamma(2-2\varepsilon)}\nn 
&=& -\frac{1}{Q^2} \left[-\frac{1}{\varepsilon} +\ln \frac{Q^2}{\mu^2}-2\right]. 
\end{eqnarray}
At the end of the day, one thus obtains   
\begin{equation}
{\cal A}_0 =\sum_q \frac{e_q^2 \alpha_s }{ Q^2}xH_g (x, \Delta_\perp)  \left[-\frac{1}{\varepsilon} +\ln \frac{Q^2}{\mu^2}-2\right] , \label{dim}
\end{equation}
which can be interpreted as the contribution to the quark GPD $xH_q (x, \Delta_\perp)$ at small-$x$, see the next section and  Appendix \ref{quark}. 
The dominant contribution for the quark GPD comes from the gluon GPD in this region.

\subsection{Helicity-flip Amplitude}
Next let us consider the the DVCS amplitude with helicity flip. It is straightforward to find 
\begin{eqnarray}
&& {\cal A}_2(\Delta_\perp)\cos 2\phi_{\Delta_\perp}=-\frac{1}{2}\sum_{\lambda}\mathcal{A}^{\lambda\neq\lambda^\prime}_T (\Delta_\perp) \\
&& \qquad \qquad = \sum_q\frac{2e_q^2 N_c}{\pi}   \int_0^1 dz z(1-z)  \int \frac{d^2 r_\perp}{r_\perp}  \epsilon_qK_1(\epsilon_q r_\perp) \cos 2\phi_{r_\perp}\int d^2q_\perp e^{-iq_\perp \cdot r_\perp}  {\cal F}(q_\perp, \Delta_\perp,z)  \nn 
&& \qquad \qquad = \sum_q \frac{2e_q^2 N_c}{\pi} \int_0^1 dz z(1-z)  \int \frac{d^2 r_\perp}{r_\perp}  \epsilon_qK_1(\epsilon_q r_\perp) \cos 2\phi_{r_\perp}\int d^2q_\perp e^{-i(q_\perp +\delta_\perp )\cdot r_\perp}  F(q_\perp, \Delta_\perp). \nonumber
\end{eqnarray}
After performing the angular integrations, we can cast the above amplitude into 
\begin{eqnarray}
{\cal A}_2(\Delta_\perp)&=&
-8\pi \sum_q e_q^2 N_c \int_0^1 dz z(1-z)  \nn 
&& \qquad \times \int_0^\infty q_\perp dq_\perp \left[  H_{02}(q_\perp, \delta_\perp)  F_0(q_\perp, \Delta_\perp)+H_{20}(q_\perp, \delta_\perp)  F_\epsilon(q_\perp, \Delta_\perp)\right],  \label{H1}
\end{eqnarray}
where 
\begin{eqnarray}
H_{02}(q_\perp, \delta_\perp) &\equiv& \int_0^\infty dr_\perp \epsilon_qK_1(\epsilon_q r_\perp)  J_{0}(q_\perp r_\perp)J_{2}(\delta_\perp r_\perp),\\
H_{20}(q_\perp, \delta_\perp) &\equiv& \int_0^\infty dr_\perp \epsilon_qK_1(\epsilon_q r_\perp)  J_{2}(q_\perp r_\perp)\left[J_{0}(\delta_\perp r_\perp)+J_{4}(\delta_\perp r_\perp)\right]. \label{fir}
\end{eqnarray}
Again the $r_\perp$-integrals can be done  \cite{grad}, but in order to make contact with the collinear calculation, let us focus on the first term in (\ref{fir}) (the other terms are subleading in the DVCS limit $Q\gg \Delta_\perp$) and evaluate it as 
\beq
&&\int_0^\infty dr_\perp \epsilon_qK_1(\epsilon_q r_\perp)  J_{2}(q_\perp r_\perp)J_{0}(\delta_\perp r_\perp) \nonumber \\ &&
=-\int_0^\infty dr_\perp \epsilon_q K_1(\epsilon_q r_\perp) \int \frac{d\phi_{r_\perp}}{2\pi}e^{ir_\perp \cdot \delta_\perp} \int \frac{d\phi_{q_\perp}}{2\pi} e^{-iq_\perp\cdot r_\perp} \cos 2(\phi_{q_\perp}-\phi_{r_\perp}) \nonumber \\ 
&& = \int_0^\infty dr_\perp \epsilon_q K_1(\epsilon_q r_\perp) \int \frac{d\phi_{q_\perp}}{2\pi} J_2(|q_\perp-\delta_\perp|r_\perp) \cos 2(\phi_{q_\perp}-\phi_{q_\perp-\delta_\perp}) \nn
&&=\frac{1}{2} \int \frac{d\phi_{q_\perp}}{2\pi}\left(1-\frac{2\delta_\perp^2 \sin^2(\phi_{q_\perp}-\phi_{\delta_\perp})}{(q_\perp-\delta_\perp)^2}\right)\left[1-\frac{\epsilon_q^2}{(q_\perp-\delta_\perp)^2}  \ln \left(1+\frac{(q_\perp-\delta_\perp)^2}{\epsilon_q^2}\right)\right] .
\eeq
We further take the collinear limit $Q^2\gg q_\perp^2$ and arrive at
\begin{eqnarray}
{\cal A}_2(\Delta_\perp)&=&
 -\sum_q \frac{e_q^2 N_c}{Q^2} \int d^2q_\perp q_\perp^2  F_\epsilon(q_\perp, \Delta_\perp) =-\sum_q \frac{e_q^2 \alpha_s \Delta_\perp^2}{4 M^2 Q^2} xE_{Tg} (x, \Delta_\perp), \label{flip}
\end{eqnarray}
 where Eq.~(\ref{fe}) is used in the last step. This should be compared to the collinear factorization calculation by Ji-Hoodbhoy \cite{Hoodbhoy:1998vm}. 
Their result reads, in the present normalization, 
\begin{equation}
{\cal A}_2=\sum_q \frac{e_q^2 \alpha_s \Delta_\perp^2}{8\pi Q^2M^2}\,  \xi \, {\rm Im} \left[\int dx\left(\frac{1}{x-\xi+i\epsilon}+\frac{1}{x+\xi-i\epsilon}\right)E_{Tg}(x,\xi)\right] \ .
\end{equation}
Noting that $E_{Tg}(x,\xi)=-E_{Tg}(-x,\xi)$, we see that  the above two are consistent with each other.

We thus see that the helicity-flip amplitude 
is proportional to the elliptic gluon distribution. Moreover, the collinear limit can be safely taken, as there is no divergence from the remaining $z$-integration. The resulting $\cos 2\phi$ correlation should be measurable in the future experiments at the EIC. A similar  observable in quasielastic scattering $\gamma_T^*p\to p'X$ has been proposed in \cite{Zhou:2016rnt}. Since these observables are  associated with the correlation in the phase space Wigner distribution \cite{Hatta:2016dxp}, such measurements will provide a unique perspective on the
 gluon tomography in nucleons at small-$x$.

\section{Momentum space calculation and the collinear limit}
\label{mom}

In this section, we repeat the calculation of the DVCS amplitude fully in momentum space and reproduce the results in the previous section. An advantage of the momentum space calculation is that it makes the connection to the collinear factorization approach more transparent. This is particularly important for the azimuthally symmetric part ${\cal A}_0$ which, as we have already seen, contains divergence in the collinear limit. We show that this divergence 
can be interpreted as that of the quark GPD contribution to the DVCS amplitude. This is because the quark GPD 
can be calculated from the gluon GPD at small-$x$. 
When we substitute the quark GPD into the collinear formula for the DVCS 
amplitude, we are able to reproduce the result of the helicity-conserved DVCS 
amplitude in the previous section. This demonstrates the 
complete consistency of the dipole and collinear factorization approaches to 
DVCS.

In momentum space, the DVCS amplitude can be straightforwardly calculated from the right diagram in
Fig.~\ref{cm}
\begin{eqnarray}
{\cal A}_T^{\mu\nu}&=& \sum_q \frac{e_q^2N_c}{2\pi}\int dzd^2q_\perp d^2q_{1\perp} (-2){ F}_x(q_\perp,\Delta_\perp) \nn 
 && \qquad \times \frac{2(z^2+(1-z)^2){q}_{1\perp}^\mu k_\perp^\nu
-q_{1\perp}^\mu k_\perp^\nu-q_{1\perp}^\nu k_\perp^\mu+ q_{1\perp}\cdot  k_\perp g_\perp^{\mu\nu}}
{q_{1\perp}^2(k_\perp^2+\epsilon_q^2)}, 
 \label{dvcs0}
\end{eqnarray}
where 
$k_\perp=q_{1\perp}+\frac{z-\bar z}{2}\Delta_\perp-q_\perp$ with $\bar{z}\equiv 1-z$  
and ${ F}_x$ is defined as in (\ref{f}). We have included a factor $-2$ to adjust to the normalization ${\cal A}_T\sim -2S$ in (\ref{As}).  
If we change variables as $\tilde{q}_\perp=q_\perp+\delta_\perp$, (\ref{dvcs0}) takes the form
\begin{eqnarray}
{\cal A}_T^{\mu\nu}&=& \sum_q \frac{e_q^2N_c}{2\pi}\int dzd^2\tilde{q}_\perp d^2q_{1\perp} (-2){\cal F}_x(\tilde{q}_\perp,\Delta_\perp,z) \nn 
&& \qquad \times \frac{2(z^2+(1-z)^2){q}_{1\perp}^\mu k_\perp^\nu
-q_{1\perp}^\mu k_\perp^\nu-q_{1\perp}^\nu k_\perp^\mu+ q_{1\perp}\cdot  k_\perp g_\perp^{\mu\nu}}
{q_{1\perp}^2(k_\perp^2+\epsilon_q^2)},
\end{eqnarray}
where $k_\perp = q_{1\perp}-\tilde{q}_\perp$ and Eq.~(\ref{com}) is used. We thus see that this shift of loop momentum is related to the appearance of the phase factor $e^{-i\delta_\perp\cdot r_\perp}$ in coordinate space discussed in Section~\ref{2}. 
For the components (\ref{co}), we obtain 
\begin{eqnarray}
{\cal A}_0=\frac{g_{\perp\mu\nu}{\cal A}_T^{\mu\nu}}{2}=-\sum_q\frac{e_q^2 N_c}{\pi}\int dzd^2q_\perp d^2q_{1\perp} \frac{(z^2+(1-z)^2){q}_{1\perp}\cdot k_\perp}
{q_{1\perp}^2(k_\perp^2+\epsilon_q^2)} {F}_x(q_\perp,\Delta_\perp) , \label{a00}
\end{eqnarray}
and 
\begin{eqnarray}
{\cal A}_2&=& \frac{h_{\perp\mu\nu}{\cal A}_T^{\mu\nu}}{2} \nn 
&=&\sum_q\frac{2e_q^2 N_c}{\pi}\int dzd^2q_\perp d^2q_{1\perp} \frac{z(1-z)\left[2
{q}_{1\perp}\cdot \Delta_\perp k_\perp\cdot \Delta_\perp-{q}_{1\perp}\cdot k_\perp \Delta_\perp^2\right]}
{q_{1\perp}^2(k_\perp^2+\epsilon_q^2)\Delta_\perp^2} 
{F}_x(q_\perp,\Delta_\perp). 
\end{eqnarray}
It is interesting to notice that the ${\cal A}_2$ depends on $\cos(2\phi)$. 
For example, we can rewrite as 
\begin{equation}
\frac{\left[2{q}_{1\perp}\cdot \Delta_\perp k_\perp\cdot \Delta_\perp-{q}_{1\perp}\cdot k_\perp \Delta_\perp^2\right]}
{\Delta_\perp^2} =q_{1\perp}k_\perp\cos(\phi_{q\Delta}+\phi_{k\Delta}) \ ,
\end{equation}
where $\phi_{q\Delta}$ and $\phi_{k\Delta}$ are azimuthal angles for $q_{1\perp}$
and $k_\perp$, respect to $\Delta_\perp$. 
 To carry out the above integrals, we define 
\begin{eqnarray}
\Gamma^{\mu\nu}(q_\perp,\Delta_\perp)=\int d^2q_{1\perp}\frac{q_{1\perp}^\mu k_{\perp}^\nu}{q_{1\perp}^2(k_\perp^2+\epsilon_q^2)} =\Gamma_0g_\perp^{\mu\nu}+\Gamma_2\tilde{q}_\perp^\mu\tilde{q}_\perp^\nu \ .
\end{eqnarray}
${\cal A}_2$ receives a contribution only from $\Gamma_2$, whereas ${\cal A}_0$ receives
from both terms.
After applying the Feynman parametrization and performing the loop integral,
$\Gamma_2$ can be written as,
\begin{equation}
\Gamma_2=-\pi \int_0^1 d\alpha\frac{\alpha}{\alpha \tilde{q}_\perp^2+\epsilon_q^2} \ .
\end{equation}
Substituting the above result into ${\cal A}_2$, we obtain
\begin{eqnarray}
 {\cal A}_2&=&-2\sum_q e_q^2 N_c \int dzd\alpha d^2q_\perp \frac{z(1-z)\alpha}{\alpha\tilde{q}_\perp^2+\epsilon_q^2}\frac{2(\tilde{q}_\perp\cdot \Delta_\perp)^2-\tilde{q}_\perp^2\Delta_\perp^2}{\Delta_\perp^2} {F}_x(q_\perp,\Delta_\perp)\ . \label{eq}
 \end{eqnarray}
By construction, (\ref{eq}) should be equivalent to (\ref{H1}), although it is difficult to see this analytically. We have checked this numerically for both the $F_0$ and $F_\epsilon$ terms.    
In the DVCS limit $\Delta_\perp\ll Q$, we can write 
\begin{equation}
\frac{2(\tilde{q}_\perp\cdot \Delta_\perp)^2-\tilde{q}_\perp^2\Delta_\perp^2}{\Delta_\perp^2}\approx q_\perp^2\cos(2\phi_{q\Delta}) \ ,
\end{equation}
 and therefore, 
\begin{eqnarray}
{\cal A}_2=-\sum_q \frac{e_q^2N_c}{Q^2} \int d^2q_\perp q_\perp^2 {F}_\epsilon(q_\perp,\Delta_\perp)=-\frac{e_q^2 \alpha_s \Delta_\perp^2}{4Q^2M^2}E_{Tg}(x,\Delta_\perp)\ ,
\end{eqnarray}
which is in agreement with (\ref{flip}). 

We now return to ${\cal A}_0$ in (\ref{a00}) and take the DVCS limit  
\begin{eqnarray}
{\cal A}_0&=&-\sum_q \frac{e_q^2 N_c}{\pi} \int{dzd^2q_\perp d^2k_{\perp}}\frac{(z^2+(1-z)^2){k}_{\perp}\cdot (k_\perp+q_\perp)}
{(k_{\perp}+q_\perp)^2(k_\perp^2+\epsilon_q^2)} {F}_x(q_\perp,\Delta_\perp)\ .
\end{eqnarray}
In order to see the infrared behavior of the above  integration more clearly, we  examine the low transverse momentum region $Q\gg k_\perp\sim q_{\perp}$ of the above integrand.  
 We first notice that only the end points of the $z$-integral 
contribute. For example, if $z\neq 1$ or $0$ so that $\epsilon_q^2\sim Q^2\gg k^2_\perp$, we immediately
find that the above integral vanishes. Therefore, we have to separate out the dominant kinematic
region of the above integration. To do that, we follow the trick of Ref.~\cite{Marquet:2009ca} and insert an identity:
$\int dx \delta\left(x-{1}/{(1+{\Lambda^2}/{\epsilon_q^2})}\right)=1$
where $\Lambda^2=(1-z)k_\perp^2+z(k_\perp+q_\perp)^2$. 
In the region  $Q\gg k_\perp\sim q_{\perp}$,  we can expand the $\delta$-function  as
\begin{eqnarray}
\delta\left(x-\frac{1}{1+\frac{\Lambda^2}{\epsilon_q^2}}\right)& =&\frac{1-z}{x}\delta\left((1-z)(1-x)-\frac{x}{z}\frac{\Lambda^2}{Q^2}\right) \nonumber\\
&=&\frac{1-z}{x}\left[\frac{\delta(1-z)}{1-x}+\frac{\delta(1-x)}{1-z}+\delta(1-x)\delta(1-z)\ln\left(\frac{Q^2}{k_\perp^2}\right)\right] \ .
\end{eqnarray}
Let us show that only the first term contributes to ${\cal A}_0$ in the above expansion. 
For that purpose, we replace $Q^2$ and $\epsilon_q^2$ by applying 
the above $\delta$-function 
$\epsilon_q^2=\frac{x}{1-x}\Lambda^2$, $Q^2=\frac{x}{z(1-z)(1-x)}\Lambda^2$ and obtain
\begin{eqnarray}
{\cal A}_0&=&-\sum_q \frac{e_q^2N_c}{\pi Q^2}\int {dxdzd^2q_\perp d^2k_{\perp}}(z^2+(1-z)^2)\frac{\Lambda^2}{z(k_\perp+q_\perp)^2}
\frac{{k}_{\perp}\cdot (k_\perp+q_\perp)}
{(1-x)k_\perp^2+x\Lambda^2}\nonumber\\
&&\times  \left[\frac{\delta(1-z)}{1-x}+\frac{\delta(1-x)}{1-z}+\delta(1-x)\delta(1-z)\ln\left(\frac{Q^2}{k_\perp^2}\right)\right] {F}_x(q_\perp,\Delta_\perp)\ .
\end{eqnarray} 
First, we can easily check that the $\delta(1-x)\delta(1-z)$ term vanishes.  Second, the term proportional to $\delta(1-x)$ also vanishes because  the integrand can be
simplified as
\begin{equation}
\frac{k_\perp\cdot (k_\perp+q_\perp)}{z(1-z)(k_\perp+q_\perp)^2} \ ,
\end{equation}
and the azimuthal integration gives zero. Thus the final result comes from the $\delta(1-z)$ term
\begin{eqnarray}
{\cal A}_0&=&-\sum_q \frac{e_q^2 N_c}{\pi Q^2}\int{dxd^2q_\perp d^2k_{\perp}}\frac{1}{1-x}\frac{{k}_{\perp}\cdot (k_\perp+q_\perp)}
{(1-x)k_\perp^2+x (k_\perp+q_\perp)^2} {F}_x(q_\perp,\Delta_\perp)\nonumber\\
&=&-\sum_q \frac{4\pi e_q^2N_c}{ Q^2}\int\frac{dxd^2q_\perp d^2k_{\perp}'}{(2\pi)^2}\frac{1}{1-x}\frac{({k}_{\perp}')^2-x(1-x)q_\perp^2}
{({k}_{\perp}')^2+x(1-x)q_\perp^2} {F}_x(q_\perp,\Delta_\perp)\ . \label{10}
\end{eqnarray}
In the collinear limit, we can further simplify this as  
\begin{eqnarray}
{\cal A}_0&=&\sum_q \frac{4\pi e_q^2N_c}{ Q^2}\int \frac{d^2k_{\perp}'}{(2\pi)^2} \frac{1}{ k'^2_{\perp}} \int d^2q_\perp q_\perp^2{F}_x(q_\perp,\Delta_\perp)\nonumber\\
&=& \sum_q \frac{2\pi e_q^2 \alpha_s}{Q^2}\int\frac{ d^2k_{\perp}'}{(2\pi)^2}\frac{1}{k'^2_{\perp}} xH_g(x)\ . \label{fin}
\end{eqnarray}
In the above calculation 
we picked up  the leading contribution in the region of $z\sim 1$, which is similar to the current fragmentation
contribution in semi-inclusive DIS at small-$x$ studied in~\cite{Marquet:2009ca}. For the 
$z\sim 0$ region, we can repeat the same procedure with $z\leftrightarrow 1-z$. 
As a result, (\ref{10}) and (\ref{fin}) are doubled and the  divergent part of the latter agrees with 
(\ref{dim}). 
In Appendix A, we show that  (\ref{fin}) can be interpreted as the quark GPD at small-$x$.

\section{Longitudinally polarized virtual photon}
\label{long}

Finally, we study the contribution from the longitudinally ($L$) polarized photon. The transition amplitude from the longitudinally polarized virtual photon to the transversely polarized real photon $\gamma^*_L p\to \gamma p'$ is usually neglected in the dipole framework and actually vanishes unless one includes the phase factor $e^{-i\delta_\perp\cdot r_\perp}$  \cite{Bartels:2003yj}. Here we calculate its contribution to the DVCS cross section. The interference term between the transverse and longitudinal virtual photon amplitudes reads 

\beq
&&\frac{L_{\mu\nu}{\cal M}^{\mu\nu}_{TL}}{W^4}=  -2{\rm Re}\sum_\lambda L^{\mu\nu}\epsilon_{\mu}^{T(\lambda)*} \epsilon_{\mu'}^{T(\lambda)} g_{\perp \alpha\beta}   {\cal A}_T^{\mu'\alpha} 
({\cal A}_L^{\nu'\beta})^* \epsilon^L_{\nu'} \epsilon_\nu^L.
\eeq
Writing $\epsilon^L_{\nu'}{\cal A}_L^{\nu'\beta}= \frac{\Delta_\perp^{\beta}}{|\Delta_\perp|} {\cal A}_L$ and using  
\beq
L^{\mu\nu}\epsilon_{\mu}^{T(\lambda)*}\epsilon_\nu^L = -\frac{2(2-y)Q}{y}\, l_\perp \cdot \epsilon_\perp^{(\lambda)*},
\eeq 
we obtain
\beq
 \frac{L_{\mu\nu}{\cal M}^{\mu\nu}_{TL}}{W^4}  &=& \frac{4(2-y)}{y}Q\,  ({\cal A}_0 + {\cal A}_2){\cal A}_L  \sum_{\lambda}   l_\perp \cdot \epsilon_\perp^{(\lambda)*} \frac{\Delta_\perp \cdot \epsilon_\perp^{(\lambda)}}{|\Delta_\perp|}  \nn 
&=& \frac{4(2-y)\sqrt{1-y}}{y^2}   Q^2\,   ({\cal A}_0 + {\cal A}_2){\cal A}_L \cos \phi_{\Delta l}.
\eeq
We immediately recognize the $\cos \phi_{\Delta l}$ angular distribution. 
${\cal A}_L$ can be evaluated as 
\beq
{\cal A}_L  &=&-\sum_q \frac{2ie_q^2 N_c Q}{\pi|\Delta_\perp|} \int_0^1 dz z (1-z)(1-2z) \nn 
&& \qquad \times \int d^2r_\perp K_0(\epsilon_q r_\perp) \frac{ r_\perp \cdot \Delta_\perp}{r_\perp^2} \int d^2q_\perp e^{-i(q_\perp + \delta_\perp) \cdot r_\perp} F(q_\perp,\Delta_\perp). 
\eeq
Naively, the $z$-integral vanishes because the integrand seems to be antisymmetric under $z\to 1-z$. However, the phase $e^{-i\delta_\perp \cdot r_\perp}=e^{-i\frac{1-2z}{2}\Delta_\perp \cdot r_\perp}$  also depends on $z$, and this makes the integral finite. Performing angular integrations, 
we find
\beq
 {\cal A}_L  &=& -\sum_q  8\pi e_q^2 N_cQ \int_0^1 dz z (1-z)(1-2z) \int_0^\infty dr_\perp K_0(\epsilon_q r_\perp) \int_0^\infty dq_\perp q_\perp \nn 
&&  \quad \times \Bigl( J_1(\delta_\perp r_\perp) J_0(q_\perp r_\perp) F_0(q_\perp,\Delta_\perp) -(J_1(\delta_\perp r_\perp)-J_3(\delta_\perp r_\perp))J_2(q_\perp r_\perp)F_\epsilon(q_\perp,\Delta_\perp) \Bigr) , 
\eeq
 where $\delta_\perp=\frac{1-2z}{2}|\Delta_\perp|$ in the argument of the Bessel functions.  
Let us ignore the $J_3(\delta_\perp r_\perp)$ term and expand as $J_1(\delta_\perp r_\perp)\approx \frac{1}{2} \delta_\perp r_\perp$. We then get a nonzero result
\beq
&& {\cal A}_L\approx -\sum_q e_q^2 N_c Q|\Delta_\perp| \int_0^1 dz z(1-z)(1-2z)^2 \int d^2q_\perp \nn 
&& \qquad \quad \times \left[ \frac{F_0(q_\perp,\Delta_\perp)}{\epsilon_q^2 + q_\perp^2}+F_\epsilon(q_\perp,\Delta_\perp) \left(\frac{1}{\epsilon_q^2+q_\perp^2}-\frac{1}{q_\perp^2} \ln \left(1+\frac{q_\perp^2}{\epsilon_q^2}\right) \right) \right]. \label{non}
\eeq 
If we do the collinear expansion, the $F_0$ term gives $xH_g(\Delta_\perp)$ via (\ref{f0}), but again the $z$-integral diverges at $z=0,1$. 
Similarly, the $F_\epsilon$ term gives $xE_{Tg}(\Delta_\perp)$ with a divergent coefficient. 
Regularizing this divergence as in (\ref{dr}), we find
\beq
{\cal A}_L = -\sum_q \frac{ e_q^2 \alpha_s |\Delta_\perp|}{ Q^3}  \left(x H_g(\Delta_\perp) + \frac{\Delta_\perp^2}{4M^2}x E_{Tg}(\Delta_\perp) \right) \frac{1}{\varepsilon} + \cdots \,. \label{al}
\eeq
The first term in (\ref{al})  again comes from the quark GPD whose contribution to the $\cos \phi$ part of the cross section is manifest in the collinear calculation (see the function called ${\cal F}^{\rm eff}$ in \cite{Belitsky:2001ns}). We are however unsure of the origin of the second term.  Presumably this arises from the twist-three part of ${\cal F}^{\rm eff}$, but we have not been able to show this explicitly. In any case, this divergence is an artifact of the collinear expansion. At the level of (\ref{non}), ${\cal A}_L$ is finite and can be used in practical calculations.

For completeness, we also note the result for the longitudinal amplitude squared 
\beq
\frac{L_{\mu\nu}{\cal M}^{\mu\nu}_{LL}}{W^4}&=& L^{\mu\nu}\epsilon_{\mu}^{L*} \epsilon_{\mu'}^{L} g_{\perp \alpha\beta}   {\cal A}_L^{\mu'\alpha} 
({\cal A}_L^{\nu'\beta})^* \epsilon^L_{\nu'} \epsilon_\nu^L = \frac{4(1-y)}{y^2}Q^2  {\cal A}_L^2. \label{last}
\eeq
Adding all the components, we arrive at the complete DVCS cross section in the dipole framework
\begin{eqnarray}
\frac{d\sigma (ep \to e'\gamma p')}{dx_BdQ^2d^2\Delta_\perp}=\frac{\alpha_{em}^3 }{\pi x_{Bj}Q^2}\Biggl\{
\left(1-y+\frac{y^2}{2}\right)({\cal A}_0^2+{\cal A}_2^2)+2(1-y) {\cal A}_0{\cal A}_2 \cos(2\phi_{\Delta l}) \nn
+(2-y)\sqrt{1-y} ({\cal A}_0+{\cal A}_2){\cal A}_L \cos \phi_{\Delta l} +(1-y){\cal A}_L^2   \Biggr\}\ .
\end{eqnarray}

\section{Conclusion}

In summary, we have studied the DVCS amplitudes at small-$x$ in the dipole formalism. The final formula for the cross section  (\ref{last}) involves the $\cos \phi$ and $\cos 2\phi$ azimuthal angular correlations. While such correlations are known in the standard collinear approach to DVCS \cite{Diehl:2003ny,Belitsky:2005qn}, it is nontrivial to retrieve them in the dipole framework. In order to obtain the $\cos \phi$ term, we have to include the (correct) phase factor $e^{-i\frac{1-2z}{2}\Delta_\perp \cdot r_\perp}$ in the amplitude. As for the $\cos 2\phi$ term, it is essential to consider the elliptic gluon Wigner distribution  \cite{Hatta:2016dxp,Hagiwara:2016kam,Zhou:2016rnt} which represents the dominant angular dependence of the dipole S-matrix.   In this regard, it is interesting to note that the elliptic gluon distribution has been recently proposed \cite{Hagiwara:2017ofm}  as a possible underlying mechanism for the observed elliptic flow ($\cos 2\phi$ azimuthal correlation among the final state hadrons) in high energy $pp$ and $pA$ collisions~\cite{Dusling:2015gta}. Thus the same distribution  
plays an important role to generate the $\cos 2\phi$ distribution  both in  
DVCS  and in inclusive hadron production in $pA$ collisions (see also \cite{Zhou:2016rnt}).  
Experimental investigations of these novel phenomena will provide 
crucial information about the 
gluon tomography in the nucleon at small-$x$.

We have also shown that, in the collinear limit, the dipole formalism reproduces the results obtained in the collinear
factorization approach for both the angular symmetric and elliptic amplitudes. As $Q^2$ is lowered, the DVCS amplitudes are sensitive to the transverse momentum distribution in the target and the dipole-CGC framework becomes more appropriate.

At last, we notice that the calculation on DVCS presented in this paper can be easily generalized to diffractive vector meson ($J/\psi$, $\rho$ and $\phi$) productions in DIS ($\gamma^\ast +p \to V +p^\prime$) (see e.g. Refs.~\cite{Brodsky:1994kf, Accardi:2012qut, Kowalski:2006hc, Kowalski:2003hm, Lappi:2014foa} and references therein), if we replace the transverse wave-function of the final state real photon by the vector meson wave-function. Similar conclusions can be also applied to this process.

\acknowledgements
This material is based upon work supported by the LDRD program of 
Lawrence Berkeley National Laboratory, the U.S. Department of Energy, 
Office of Science, Office of Nuclear Physics, under contract number 
DE-AC02-05CH11231 and by the Natural Science Foundation of China (NSFC) under Grant No.~11575070.

\appendix

\section{Collinear Factorization Results and Quark GPD and PDF at Small-$x$ }
\label{quark}

The DVCS amplitude is calculated in terms of the off-forward tensor $T^{\mu\nu}$,
\begin{equation}
T^{\mu\nu}=i\int d^4z e^{-iq\cdot z}\langle P'|j^\mu(z/2)j^\nu(-z/2)|P\rangle \equiv g_\perp^{\mu\nu}T_0+h_\perp^{\mu\nu} T_2  \ .
\end{equation}
The above two terms have been calculated in the literature. In small-$x$ limit, they 
take the following forms~\cite{Ji:1996ek,Hoodbhoy:1998vm},
\begin{eqnarray}
T_0&=&- \sum_q e_q^2\int dx \,  \alpha(x) H_q(x,\xi,\Delta_\perp^2) \ ,\\
T_2&=& \sum_q e_q^2\frac{\alpha_s}{4\pi}\frac{\Delta_\perp^2}{4M^2}\int dx \, \alpha (x) E_{Tg}(x,\xi,\Delta_\perp^2) \ ,
\end{eqnarray}
where $H_q$ and $E_{Tg}$ are the quark GPD and helicity-flip gluon GPD, and $\alpha(x)$ is defined as
\begin{equation}
\alpha(x)=\frac{1}{x-\xi+i\epsilon}+\frac{1}{x+\xi-i\epsilon} \ .
\end{equation}
The other contribution in $T_2$ is suppressed at small-$x$, and has been neglected in the
above. We are particularly interested in the imaginary part of the scattering amplitudes
\begin{eqnarray}
{\rm  Im}\, T_0&=&\frac{\pi}{\xi}\sum_q e_q^2\left[\xi H_q(\xi,\xi,\Delta_\perp^2)+\xi H_{\bar q}(\xi,\xi,\Delta_\perp^2)\right] \ ,\\
{\rm Im}\, T_2&=&-\frac{\pi}{\xi}\frac{\alpha_s}{2\pi}\frac{\Delta_\perp^2}{4M^2}\sum_q e_q^2\xi E_{Tg}(\xi,\xi,\Delta_\perp^2)  \ ,
\end{eqnarray}
where we have taken into account the antiquark contribution, $H_q(-x,\xi,\Delta_\perp^2)=-H_{\bar q}(x,\xi,\Delta_\perp^2)$.

At small-$x$, the quark distribution comes from the gluon splitting. The forward quark 
distribution can be calculated as
\begin{equation}
xq(x)=\frac{\alpha_s}{2\pi}\frac{1}{2}\int_x^1 d\zeta\bigl( \zeta^2+(1-\zeta)^2\bigr) x'G(x')\int\frac{ dk_{\perp}^2}{k_{\perp}^2} \ ,
\end{equation}
where $\zeta=x/x'$ and $G(x')$ is the integrated forward gluon distirbution. By applying the small-$x$ approximation, the above can be
simplified as
\begin{equation}
xq(x)\approx xG(x) \frac{\alpha_s}{2\pi}\frac{1}{2}\cdot \frac{2}{3} \int\frac{ dk_{\perp}^2}{k_{\perp}^2} \ ,\label{equarkx}
\end{equation}
 where we assumed that $x'G(x')$ is approximately constant at small-$x'$. 
For the quark GPD, the evolution equation depends on the
skewness parameter $\xi$ which reads, for $x>\xi$, 
\begin{equation}
xH_q(x,\xi,\Delta_\perp^2)=\frac{\alpha_s}{2\pi}\frac{1}{2}\int_x^1 d\zeta\frac{\zeta^2 + (1-\zeta)^2-\frac{\xi^2}{x^2}\zeta^2}{(1-\frac{\xi^2}{x^2}\zeta^2)^2}x'H_g(x',\xi,\Delta_\perp^2)\int\frac{ dk_{\perp}^2}{k_{\perp}^2} \ ,
\end{equation}
where $H_g(x',\xi,\Delta_\perp^2)$ is the gluon GPD. The limit $x\to \xi$ requires some care because of the singularity. If one naively sets $\xi=x$ in the integrand and assumes that $x'H_g(x',\xi)$ is a constant, the $\zeta$-integral gives 
$\int_x^1  \frac{d\zeta}{(1+\zeta)^2} \approx \frac{1}{2}$. However, this is incorrect. One has to first evaluate the $\zeta$-integral exactly and then take the limit $x\to \xi$. This gives 
\beq
\lim_{x\to \xi} \int_x^1 d\zeta\frac{\zeta^2 + (1-\zeta)^2-\frac{\xi^2}{x^2}\zeta^2}{(1-\frac{\xi^2}{x^2}\zeta^2)^2} = \frac{1}{1+\xi}\approx 1.
\eeq  
We thus find 
\begin{equation}
\xi H_q(\xi,\xi)\approx \xi H_g(\xi,\xi) \frac{\alpha_s}{2\pi}\frac{1}{2}\cdot 1 \int\frac{ dk_{\perp}^2}{k_{\perp}^2}\  .
\end{equation}
It is interesting to notice that here the prefactor is 1, instead of $\frac{2}{3}$
for the forward quark distribution in Eq.~(\ref{equarkx}). Substituting the above results, we obtain the collinear factorization
result for the DVCS amplitudes at small-$x$,
\begin{eqnarray}
{\rm Im} \, T_0&=&\frac{\alpha_s}{2\xi }\sum_q e_q^2\xi H_g(\xi,\xi,\Delta_\perp^2)\int\frac{ dk_{\perp}^2}{k_{\perp}^2}   \label{A1} \ ,\\
{\rm Im}\, T_2&=&-\frac{\alpha_s}{2\xi }\frac{\Delta_\perp^2}{4M^2}\sum_q e_q^2\xi E_{Tg}(\xi,\xi,\Delta_\perp^2)   \label{A2}\ ,
\end{eqnarray}
where we have combined the quark and antiquark contributions together. To compare to
our results in this paper, we note that the normalizations for the
hadronic tensor are different,
\begin{equation}
{\rm Im} \, T^{\mu\nu}=W^2A^{\mu\nu}=\frac{Q^2}{x_{Bj}}A^{\mu\nu}\approx \frac{Q^2}{2\xi}A^{\mu\nu} \ .
\end{equation}
We thus find that (\ref{A1}) agrees with (\ref{dim}) or (\ref{fin}) (the latter has to be multiplied by 2 as noted above (\ref{fin})), and (\ref{A2}) agrees with  (\ref{flip}).

\end{document}